\documentclass[
preprint,
 amsmath,amssymb,
 aps,
prb,
]{revtex4}

\usepackage{graphicx}
\usepackage{dcolumn}
\usepackage{bm}
\usepackage{hyperref}
\usepackage{multirow}

\begin{document}


\title{Multipolar expansions for scattering and optical force calculations beyond the long wavelength approximation}

\author{Marco Riccardi, Andrei Kiselev, Karim Achouri and Olivier J.~F.~Martin}
\email{olivier.martin@epfl.ch}
\affiliation{Nanophotonics and Metrology Laboratory, Swiss Federal Institute of Technology Lausanne (EPFL), EPFL-STI-NAM, Station 11, CH- 1015 Lausanne, Switzerland}

\begin{abstract}	
We review three different approaches for the calculation of electromagnetic multipoles, namely the Cartesian primitive multipoles, the Cartesian irreducible multipoles and the spherical multipoles. We identify the latter as the best suited to describe the scattering of electromagnetic radiation, as exemplified for  an amorphous silicon sphere. These multipoles are then used to calculate the optical force acting on semiconductor, dielectric or metallic particles in a wide wavelength range, from the dipolar down to the Mie regimes.
\end{abstract}

\maketitle

\section{Introduction}

Light-matter interactions empower a plethora of natural phenomena that represent the foundations of a broad variety of fields such as quantum optics and technologies \cite{Mandel1995, Wieman1999}, dynamic light scattering \cite{Koman2015, Stetefeld2016}, microscopy \cite{Hillenbrand2002, Taylor2019} and integrated optics \cite{Stegeman1985, Hunsperger2009}, to name a few. Unfortunately, a full characterization of these interactions leads to the formulation of complex field equations whose analytical solutions are known only for a few simple cases, notably for the scattering by a spherical particle \cite{VandeHulst1957, Bohren1983}. To facilitate the analysis of more complex systems, a multipolar expansion can be employed to gain physical insights into the underlying light-matter interactions. The strength of this approach lies in the fact that, by considering only the first few leading terms of the expansion, a sufficient convergence of the multipolar solution can be achieved without the need to undertake the full calculation, thus greatly simplifying the problem. For this reason, multipolar expansions have been exploited for a broad variety of applications including the study and design of radiation patterns for electromagnetic sources \cite{Ziolkowski2017, Shen2020}, the determination of molecular and atomic polarizabilities \cite{Mclean1967, Mitroy2010, Kern2012} and the design and characterization of metasurfaces \cite{Muhlig2011, Yang2017a, Dezert2019, Terekhov2019, Achouri2021, Achouri2022}. Other examples include the scattering of electromagnetic radiation by an object \cite{Lank2018, Alaee2018, Gurvitz2019, Kiselev2019, Ray2021} and the generation of optical forces on a scatterer \cite{Salandrino2012, Jiang2015, Zhang2015, Lank2018, Achouri2020, Kiselev2020}. Surprisingly, despite their success and widespread use, the definition of multipoles is still ambiguous in the literature, where different strategies to perform the multipolar expansion have been reported \cite{Zhang2015, Fernandez-Corbaton2017, Alaee2018, Li2018, Gurvitz2019} and considerable confusion persists over what formulation to use when \cite{Evlyukhin2019}. In this contribution, we revisit and clarify the derivation and limitations of three different multipole formulations and apply them to the calculation of the scattering cross section and optical force for spherical particles with a broad variety of materials and illumination wavelengths, from the long wavelength down to the Mie regime.

The paper is divided into three main sections specifically dealing with the calculations of electromagnetic multipoles -- Section \ref{S_multi} -- and, subsequently, their applications to scattering cross section and optical force calculations, Section \ref{S_SCS} and Section \ref{S_force} respectively. In particular, Section \ref{S_multiP} introduces the primitive electromagnetic multipoles as arising from the Taylor expansion of the retarded vector potential -- Eq.\ (\ref{expA}). Section \ref{irrTor}, shows how to derive the irreducible and toroidal multipoles from the primitive ones, see Figure \ref{fig3}, and elucidates their physical interpretation. The spherical multipoles are then introduced in Section \ref{S_sph} as the coefficients of the decomposition of the current onto the vector spherical harmonics in momentum space -- Eq.\ (\ref{Jsp}) -- and their connection to the irreducible and toroidal moments are discussed. Section \ref{S_SCS} presents scattering cross section calculations performed with these three formulations of multipoles and compares them to the predictions of Mie theory, while Section \ref{S_force} shows the multipolar expansion, using the spherical multipoles, of the optical force acting on particles of different materials.

\section{Electromagnetic multipoles}\label{S_multi}

\begin{figure*}
\includegraphics[scale=0.5]{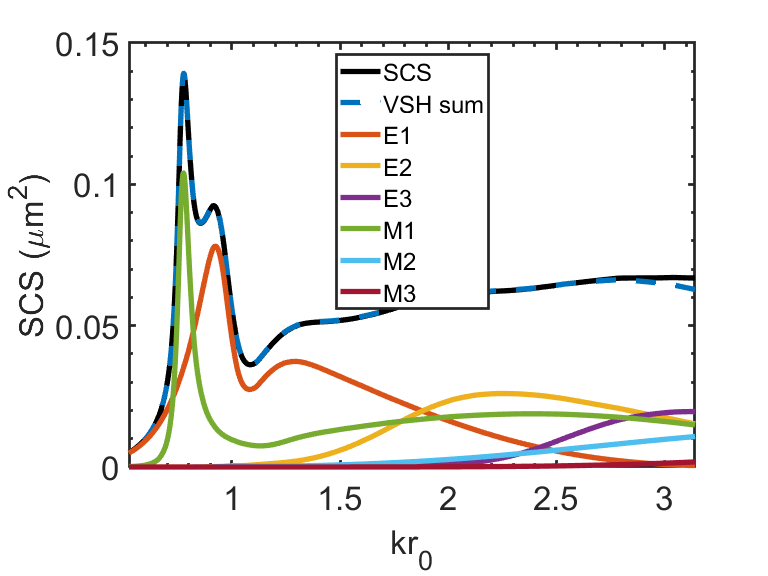}
\caption{Mie scattering cross section for an amorphous silicon sphere of radius $r_0 = 100$ nm decomposed into the first three orders of the vector spherical harmonics. E1, E2 and E3 represent the first three electric VSH; and similarly for their magnetic counterparts M1, M2 and M3. The blue dashed line represents the sum of the six VSH and agree well with the SCS.}
\label{VSH}
\end{figure*}

Let us start by considering a material volume $V$ immersed in vacuum. Under the influence of incoming harmonic electromagnetic radiation, a current density $\mathbf{J}(\mathbf{r},t)$ is produced inside the material. This current density is responsible for the generation of the scattered electric and magnetic fields, from which all other electromagnetic quantities of interest can be computed. These fields can be expressed using the retarded vector potential in the Lorentz gauge,
\begin{equation}
\mathbf{A}(\mathbf{R},t) = \frac{\mu_0}{4\pi} \int_V \frac{\mathbf{J}(\mathbf{r},t - \frac{|\mathbf{R} - \mathbf{r}|}{c})}{|\mathbf{R} - \mathbf{r}|} d\mathbf{r} \,,
\label{potential}
\end{equation}
where $\mu_0$ represents the vacuum permeability, $\mathbf{R}$ the vector to the point of observation and $\mathbf{r}$ the vector to a point in the current distribution \cite{1386}. We use the MKSA unit system and assume harmonic fields with an $e^{-i \omega t}$ time dependence throughout. As already pointed out in the introduction, this integral representation of $\mathbf{A}$ is unpractical for most applications and a multipolar expansion of the vector potential is therefore performed to study light-matter interacting systems. Different strategies can be used to realize this decomposition and, in particular, different derivations have been proposed either in Cartesian \cite{Gurvitz2019} or spherical \cite{Alaee2018} coordinates to obtain a set of tensors, called multipoles, which are used to provide an approximate description of the scattering system. In this framework, each multipole represents a specific type of light-matter interaction, whose contribution can be singled out from the total scattering response of the system and analysed thanks to the multipolar decomposition. To be able to do so we recall that, as will be shown in Section \ref{irrTor}, only those tensors that are irreducible under the SO(3) group transformations, meaning that they are both symmetric and traceless \cite{1386}, can be used to properly model a physical system. This constrain stems from the physical requirement that any physical property be conserved under rotations and parity transformations. As a consequence, we need to be able to express an electromagnetic multipole in terms of the vector spherical harmonics functions (VSH), which form the basis for any irreducible representation in SO(3). These functions are the eigenfunctions of the square of the orbital angular momentum operator and are plotted in Figure 1 in the Supporting Information. In particular, for the spherically-symmetric objects that we consider here, it must be possible to express each multipole in terms of only one order $l$ of either an electric $\mathbf{N}_{lm}(\mathbf{R})$ or magnetic $\mathbf{M}_{lm}(\mathbf{R})$ VSH defined in Eq.\ (3) and (4) in the Supporting Information \cite{Gladyshev2020}. Any multipole which does not satisfy this requirement will not provide a good description of a physical system and will not carry any real physical meaning. In the following, we will first consider a naive derivation of the Cartesian multipoles, termed \emph{primitive} multipoles, and show a strategy to improve on it by deriving a more accurate formulation using \emph{irreducible} and \emph{toroidal} multipoles. Second, we will show how to compute the \emph{spherical} multipoles, and highlight the differences between the various formulations. We will then apply these three different multipole formulations to the calculation of the scattering cross section (SCS) of an amorphous silicon sphere of radius $r_0 = 100$ nm, for which an exact expression can be found using Mie theory. To this end, Figure \ref{VSH} shows the Mie scattering cross section and its decomposition into the first six VSH.
This decomposition is performed by projecting the scattered electric far field into the electric and magnetic VSH, yielding \cite{Alaee2019}
\begin{equation}
	\mathbf{E}_\text{scat}(\mathbf{R}) = \sum_{l=1}^{\infty} \sum_{m=-l}^{m=l} a^f_{lm} \mathbf{N}_{lm}(\mathbf{R}) + b^f_{lm} \mathbf{M}_{lm}(\mathbf{R})
	\label{Miefield} \:\: .
\end{equation}
In the above, the $a^f_{lm}$ and $b^f_{lm}$ terms are the so-called \emph{Mie coefficients} and represent the projection of the scattered field onto the corresponding VSH \cite{Jackson1998}. Alternatively, the scattered far field can be decomposed into its multipolar components yielding, up to the octupolar order, \cite{Gurvitz2019}
\begin{eqnarray}
\mathbf{E}_\text{scat}(\mathbf{R}) = \frac{k^2}{4 \pi \epsilon_0} \frac{e^{i k R}}{R} \left\{ \mathbf{n} \times \mathbf{P} \times \mathbf{n} + \frac{1}{c} \mathbf{M} \times \mathbf{n} +\right.
\nonumber\\
+ \frac{ik}{2} \mathbf{n} \times \left[ \mathbf{n} \times \left(\underline{\underline{\mathbf{EQ}}} \cdot \mathbf{n} \right)\right] + \frac{ik}{2c} \mathbf{n} \times \left(\underline{\underline{\mathbf{MQ}}} \cdot \mathbf{n} \right) +
\label{multiE}\\
+ \left. \frac{k^2}{6} \mathbf{n} \times \left[ \mathbf{n} \times \left(\underline{\underline{\underline{\mathbf{EO}}}} \cdot \mathbf{n} \cdot \mathbf{n} \right)\right] + \frac{k^2}{6c} \mathbf{n} \times \left(\underline{\underline{\underline{\mathbf{MO}}}} \cdot \mathbf{n} \cdot \mathbf{n} \right) \right\}
\nonumber \:\: ,
\end{eqnarray}
where $\mathbf{P}$, $\underline{\underline{\mathbf{EQ}}}$ and $\underline{\underline{\underline{\mathbf{EO}}}}$ are the electric dipole, quadrupole and octupole moments, $\mathbf{M}$, $\underline{\underline{\mathbf{MQ}}}$ and $\underline{\underline{\underline{\mathbf{MO}}}}$ their magnetic counterparts with $R = |\mathbf{R}|$ and $\mathbf{n} = \mathbf{R} / R$. Bold quantities represent first rank tensors (vectors), double-underlined second rank tensors (matrices) and triple-underlined third rank tensors. $\epsilon_0$ is the vacuum permittivity, $c$ the speed of light in vacuum and $k = \omega / c = 2\pi / \lambda$, with $\lambda$ being the wavelength of the incoming radiation. In principle, Eq.\ (\ref{multiE}) is valid for any type of multipoles and by inserting those of a specific family (primitive, irreducible or spherical moments), we will be able to compute different multipolar cross sections and compare them to the Mie predictions given by Eq.\ (\ref{Miefield}), and shown in Figure \ref{VSH}, in order to judge what multipole formulation describes the scattering process best. The calculations are performed using a surface integral equation approach \cite{Kern2009, Gallinet2010} for an $x$-polarized plane wave propagating along the $z$-axis of the sphere. The optical properties for amorphous silicon are taken from the literature \cite{Pierce1972}.

\subsection{Primitive Cartesian multipoles}\label{S_multiP}

To derive the primitive multipoles, the current $\mathbf{J}(\mathbf{r},t - |\mathbf{R} - \mathbf{r}| / c)$ and the $1 / |\mathbf{R} - \mathbf{r}|$ term in the integrand of the expression for the vector potential $\mathbf{A}$ in Eq.\ \ref{potential} are expanded as a Taylor series for $|\mathbf{r}| \ll |\mathbf{R}|$, i.e.\ $|\mathbf{r}| \to 0$. The advantage of this method lies in the fact that, for arbitrarily small sources, only the first few terms of the Taylor series need to be considered to completely describe the system, greatly simplifying the calculations. However, as the scatterer size increases, more and more terms are needed to correctly model the interaction and their computation becomes increasingly cumbersome. For the current, this expansion yields \cite{1386}
\begin{equation}
\mathbf{J}(\mathbf{r},t - \frac{|\mathbf{R} - \mathbf{r}|}{c}) = \left. \sum_{n=0}^\infty \frac{r^n}{n!} \nabla^n \mathbf{J}(\mathbf{r},t - \frac{|\mathbf{R} - \mathbf{r}|}{c}) \right|_{\mathbf{r} = 0}
\label{expJ}
\end{equation}
and the potential, after further manipulation with the continuity equation \cite{Jackson1998}, can be rewritten as (see Sec.\ 4.1.2 in \cite{1386})
\begin{eqnarray}
\mathbf{A}(\mathbf{R},t) = \frac{\mu_0}{4\pi} \sum_{l=1}^\infty \frac{(-1)^{l-1}}{l!} \nabla^{l-1} \left[ \frac{1}{R} \underset{\cdot \cdot \, l}{\dot{\underline{\mathbf{P}}}}^P(\tau) + \nabla \times \frac{1}{R} \underset{\cdot \cdot \, l}{\underline{\mathbf{M}}}^P(\tau) \right]
\label{expA} \:\: ,
\end{eqnarray}
where the dot means a time derivative with respect to $\tau = t - R / c$. We see that two new quantities appear in the expression of $\mathbf{A}$: the primitive electric multipoles $\underset{\cdot \cdot \, l}{\underline{\mathbf{P}}}^P$ and the primitive magnetic multipoles $\underset{\cdot \cdot \, l}{\underline{\mathbf{M}}}^P$. This new notation for the multipoles indicates a tensor of rank $l$. The general definitions of these multipoles are given in Eqs.\ (6) and (7) in the Supporting Information, while Table I in the Supporting Information provides specific expressions for the first three electric and magnetic multipoles. 

\begin{figure*}
	\includegraphics{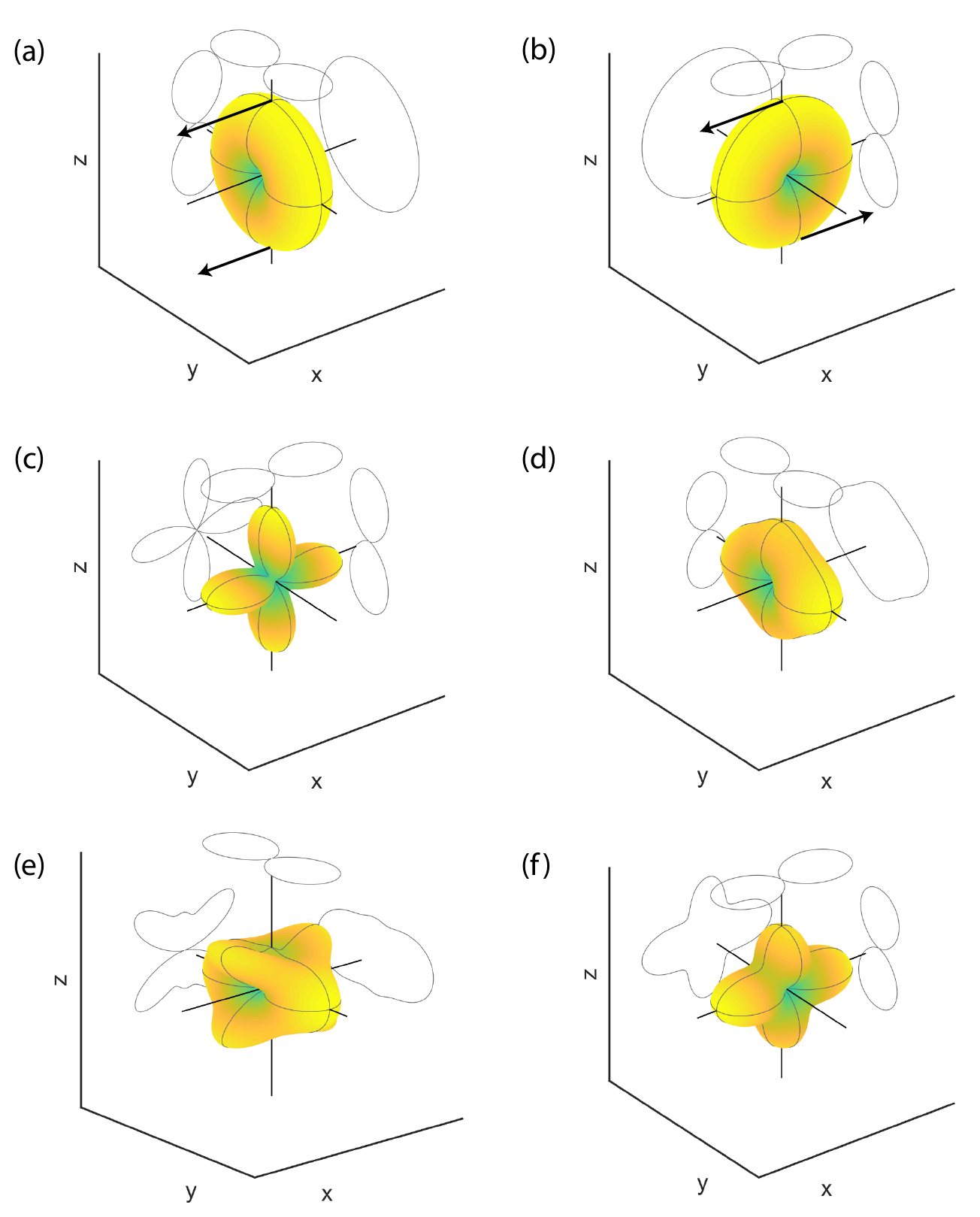}
	\caption{3D distribution of the normalized scattered intensity of the primitive electric (left) and magnetic (right) multipoles for $\lambda = 400$ nm illumination. (a)-(b) Dipole moments, where the black arrows indicate the direction of the electric field at the selected points. (c)-(d) Quadrupole moments. (e)-(f) Octupole moments.}
	\label{fig1}
\end{figure*}

Let us first point out that electric and magnetic moments of the same order $l$ arise at different orders $n$ of the Taylor expansion, in such a way that, for example, for $n = 0$ one recovers the first electric multipole, while for $n = 1$ one obtains the second electric multipole and the first magnetic one \cite{Gurvitz2019}. As a consequence, $\mathbf{P}^P$ is the first multipole arising from the Taylor series and is the leading term describing the interacting light-matter system. By inserting $\mathbf{P} = \mathbf{P}^P$ in Eq.\ (\ref{multiE}), it is possible to calculate the contribution of the first electric moment to the scattered field intensity, whose 3D radiation pattern is plotted in Figure \ref{fig1}(a).
We can clearly see here that this contribution is analogous to that of an oscillating electric dipole along the $x$ axis and described by $\mathbf{N}_{11}$ in Eq.\ (\ref{Miefield}). Therefore $\mathbf{P}^P$ takes the name of \emph{electric dipole}. Similarly, $\mathbf{M}^P$ is described by $\mathbf{M}_{1-1}$ and is called the \emph{magnetic dipole}. Interestingly however, if we consider higher order multipoles, their radiation patterns do not present the typical distributions of the corresponding $l = 2$ and $l = 3$ VSH. In particular, we see that the intensity patterns of the magnetic quadrupole and octupole seem to present characteristic radiation features of lower order multipoles such as the electric dipole and quadrupole, respectively. It is therefore clear that, while the expansions outlined in Eqs.\ (\ref{expJ}) and (\ref{expA}) are mathematically correct and allow us to derive the primitive multipoles, they do not carry any physical meaning as it is not possible to express them in terms of the corresponding VSH. The same conclusion can be reached when considering Eq.\ (6) and Table I in the Supporting Information, where it is shown that the primitive electric multipoles are not irreducible, as they are fully symmetric but, in general, not traceless. This is why the primitive electric quadrupole is not a physically appropriate tensor, even though it radiates as a proper quadrupolar VSH. The primitive magnetic multipoles, on the other hand, are neither symmetric nor traceless (see Eq.\ (7) and Table I in the Supporting Information), with the magnetic quadrupole being an exception as it is traceless but not symmetric. A proper physical description of the scattering system can therefore be achieved by detracing the primitive electric multipoles and symmetrizing and detracing the magnetic ones, in order to obtain a set of irreducible tensors able to describe our system. Before tackling this task let us point out that, since the properties of symmetry and tracelessness are not defined for first order tensors (vectors), the primitive dipole moments can already carry some definite physical meaning: they represent the response of the system when the scatterer can be modelled as a linearly oscillating pair of charges (electric dipole) or as a closed-loop circular current (magnetic dipole).

\subsection{Irreducible and toroidal Cartesian multipoles}\label{irrTor}

\begin{figure*}
\includegraphics{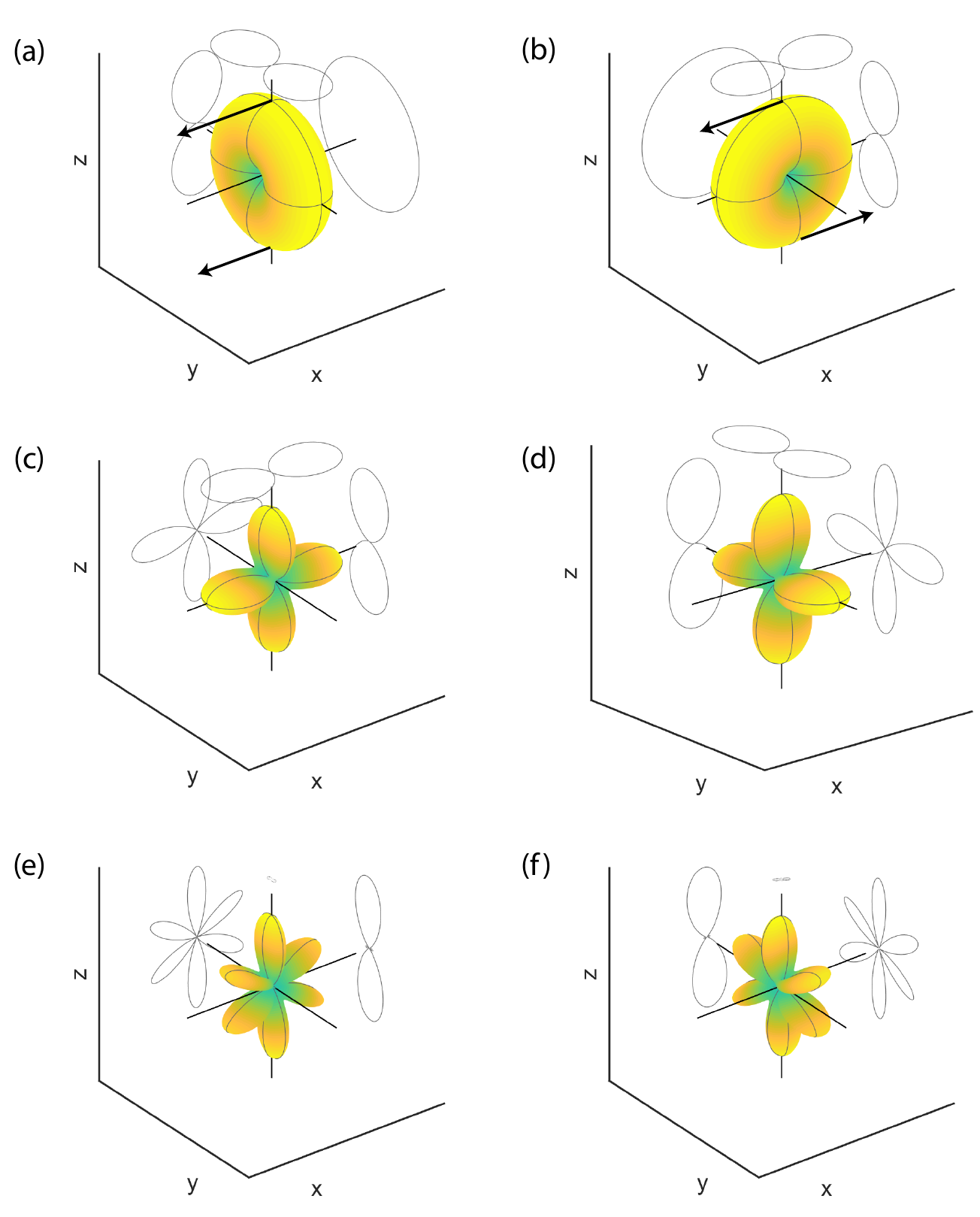}
\caption{3D distribution of the normalized scattered intensity of the irreducible electric (left) and magnetic (right) multipoles for $\lambda = 400$ nm illumination. (a)-(b) Dipole moments. (c)-(d) Quadrupole moments. (e)-(f) Octupole moments.}
\label{fig2}
\end{figure*}

The symmetrization and detracing procedure to extract the irreducible Cartesian multipoles from the primitive ones has already been described in detail elsewhere \cite{1386, Gurvitz2019} and we will not dive too much into the details. It is important to note, however, that this procedure allows the decomposition of a primitive moment $\underset{\cdot \cdot \, l}{\underline{\mathbf{A}}}^P$ into
\begin{equation}
\underset{\cdot \cdot \, l}{\underline{\mathbf{A}}}^P = \underset{\cdot \cdot \, l}{\underline{\mathbf{A}}}^I + \underset{\cdot \cdot \, p}{\underline{\mathbf{A}}}^* \:\: ,
\label{decomp}
\end{equation}
where $\underset{\cdot \cdot \, l}{\underline{\mathbf{A}}}^I$ is the irreducible representation of $\underset{\cdot \cdot \, l}{\underline{\mathbf{A}}}^P$ and $\underset{\cdot \cdot \, p}{\underline{\mathbf{A}}}^*$ is the residual left after the symmetrization / detracing procedure, which will generally have a lower rank than the primitive and irreducible moments, i.e. $p < l$. Figure \ref{fig2} shows the radiation patterns of the irreducible moments and reveals how, thanks to their irreducibility, these multipoles radiate with the expected dipolar, quadrupolar and octupolar behaviour proper of the $l = 1$, $l = 2$ and $l = 3$ VSH.
As for the primitive multipoles, we also provide in Table II in the Supporting Information the definition of the first few irreducible moments. We note that, as already mentioned, the irreducible electric and magnetic \emph{dipoles} are the same as the primitive ones.

\begin{figure*}
\includegraphics[width=0.95\textwidth]{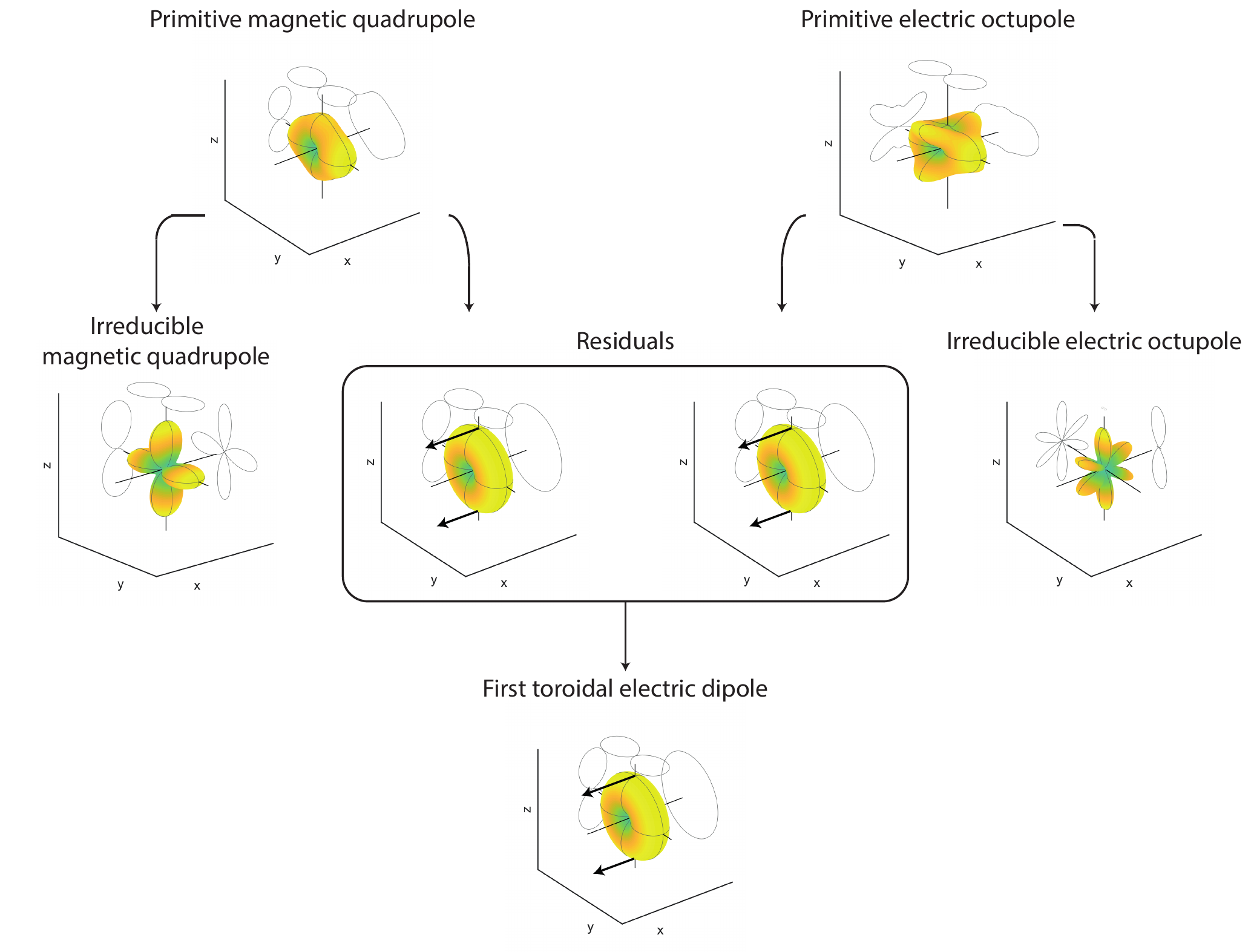}
\caption{Normalized intensity radiation patterns showing the decomposition of the primitive magnetic quadrupole and primitive electric octupole, arising at the second order of the Taylor series, into their irreducible and residual parts, according to Eq.\ (\ref{decomp}). Their residuals are then combined to generate the first toroidal electric dipole.}
\label{fig3}
\end{figure*}

After deriving physically-valid irreducible multipoles, the question arises now as to how to deal with the residuals produced by the symmetrization / detracing procedure. Surprisingly, as shown in Figure \ref{fig3} for the $n = 2$ case, it is found that these tensors have well-defined far field distributions analogous to those of the irreducible multipoles. In particular, when detracing and symmetrizing a primitive multipole of rank $l$, its residual will radiate as a multipole of lower rank $p$, meaning that the primitive multipoles are not really “pure” moments but contain contributions from different orders of the VSH and are therefore not completely independent. We explicitly show this counter-intuitive behaviour for a traceless, but not symmetric, matrix representing a fictitious primitive magnetic quadrupole:
\begin{equation}
\underline{\underline{\mathbf{A}}}^P =
\begin{pmatrix}
	2 & 5 & 4 \\
	7 & 1 & 8 \\
	3 & 11 & -3 
\end{pmatrix} \:\: .
\end{equation}
This multipole, being non-symmetric, is non-invariant under rotation transformations and is therefore still reducible. Its irreducible form can be derived by decomposing it into its symmetric and antisymmetric parts ($I$ represents the identity matrix)
\begin{eqnarray}
\underline{\underline{\mathbf{A}}}^P = \frac{1}{2} (A^P_{ij} + A^P_{ji}) + \frac{1}{2} (A^P_{ij} - A^P_{ji}) =
\nonumber \\
= \frac{1}{2} \begin{pmatrix}
	4 & 12 & 7 \\
	12 & 2 & 19 \\
	7 & 19 & -6 
\end{pmatrix}
+ \frac{1}{2} \begin{pmatrix}
	0 & -2 & 1 \\
	2 & 0 & -3 \\
	-1 & 3 & 0 
\end{pmatrix} =
\\
= \frac{1}{2} \begin{pmatrix}
	4 & 12 & 7 \\
	12 & 2 & 19 \\
	7 & 19 & -6 
\end{pmatrix}
+ \frac{1}{2} I \times
\begin{pmatrix}
	3 \\
	1 \\
	2  
\end{pmatrix} = \underline{\underline{\mathbf{A}}}^I + \mathbf{A}^*
\nonumber
\end{eqnarray}
of which the latter, representing the residual after the symmetrization procedure, depends on only three parameters and therefore behaves like a lower rank tensor, i.e.\ a dipole, and is the first contribution to the electric toroidal dipole, as shown in Figure \ref{fig3}. The former part, on the other hand, is now  a traceless \emph{and} symmetric matrix representing the fictitious irreducible magnetic quadrupole. This explicitly shows the limitation of Cartesian multipoles: it is impossible to completely describe, for example, the far field electric dipolar response of a system by only considering the electric dipole moment: a complete and exact description of such a system needs to take into account the residuals of all the other high order multipoles, which is impractical and defies the scope of employing the multipolar expansion. However, by only considering the first few orders of the Taylor expansion, the first additional contributions to the low order multipoles can be derived \cite{Gurvitz2019}. This is done by combining the residuals left after the derivation of the irreducible moments of same order $n$, generating the toroidal multipoles, as shown in Figure \ref{fig3} for the first toroidal electric dipole $\mathbf{T}_1^P$.
This arises by a combination of the residuals of the primitive electric octupole and primitive magnetic quadrupole ($n = 2$), which radiate as pure electric dipoles and therefore provide a first correction to $\mathbf{P}^I$, which is now written as $\mathbf{P} = \mathbf{P}^I + ik \mathbf{T}_1^P /c$ \cite{Gurvitz2019}. Here, the factor $ik /c$ takes into account the phase difference between $\mathbf{P}^I$ and $\mathbf{T}_1^P$. The superscript $P$ indicates a toroidal moment of the electric kind, as opposed for example to $\mathbf{T}_1^M$ which is the first toroidal magnetic dipole. Similarly, it can be shown that the residuals of the primitive electric 32-pole and magnetic hexadecapole ($n = 4$) combine to generate the second toroidal electric dipole $\mathbf{T}_2^P$, which contributes to the total electric dipole as $\mathbf{P} = \mathbf{P}^I + ik \mathbf{T}_1^P /c + ik^3 \mathbf{T}_2^P /c$ \cite{Gurvitz2019}. Higher order multipoles also produce in principle additional electric toroidal dipole terms, together with higher order electric and magnetic toroidal moments, as shown in Table III in the Supporting Information. It is thus clear that these toroidal moments act as higher order corrections to the irreducible multipoles and are therefore specifically relevant at high energy illuminations. Physically, they represent different charge and current configurations that radiate as normal irreducible moments, from which they are therefore virtually indistinguishable in the far field \cite{Papasimakis2016}. We point out that a complete agreement in the literature as to whether the toroidal multipoles really represent a new independent multipole family, together with the irreducible electric and magnetic ones, has not yet been reached \cite{1386, Papasimakis2016, Fernandez-Corbaton2017, Talebi2018}, and we will come back to this issue in the next section. Regardless of this fact, they represent real non trivial charge and current configurations in the source which contribute to the total far field emission, and need therefore to be considered for a full representation of the scattering response \footnote{Let us point out that a complete description of a radiating source needs to also consider the mean square radii of the multipoles, which describe charge and current configurations with non trivial radial distributions but similar radiation pattern as their parent multipoles \cite{Nemkov2018}.}.

We have now reviewed a physically appropriate definition of the electromagnetic multipoles, where each moment is not just simply a term in the Taylor series of the potential but carries some clear and definite physical meaning, and can therefore be used to properly model field-matter interactions. For example, these moments can be employed to interpret non-radiating charge and current distributions, termed anapole states \cite{Yang2019}, as simply arising from the destructive interference between irreducible and toroidal moments of the same kind \cite{Miroshnichenko2015, ChiehWu2018}. On the other hand, as we have already discussed, it is impractical to use this formulation to capture the full scattering response arising at a certain order, as in principle infinite toroidal corrections would need to be added to the corresponding irreducible moment. This represents a significant limitation, especially when several modes are excited in the scatterer and the first known toroidal terms do not provide an adequate correction to the irreducible response. Luckily, this shortcoming can be overcome by employing the spherical multipoles discussed in the next section.

\subsection{Spherical multipoles}\label{S_sph}

The key difference between the derivation of the Cartesian and spherical multipoles is that, as one might guess, the latter are first derived in spherical coordinates and only later converted into the Cartesian system for ease of use. An advantage of this approach lies in the fact that the as-derived spherical moments are already irreducible \cite{Jackson1998} and one does not need to go through the whole symmetrization and detracing procedure, as for the Cartesian multipoles, to obtain physically-relevant moments. Moreover, their accuracy does not depend on the scatterer size and they can therefore be used to model light-matter interactions for arbitrarily large spherically-symmetric sources.

To derive the spherical multipoles, a Fourier transform of the current is first performed to derive its energy-momentum components $\mathbf{J}_\omega(\mathbf{k})$. Of these, only those components $\overset{\circ}{\mathbf{J}}_\omega(\mathbf{k})$ defined on the spherical shell domain satisfying $|\mathbf{k}| = \omega / c$ radiate transverse electromagnetic fields outside the source \cite{Devaney1974, Fernandez-Corbaton2015} and are therefore of interest for the following derivation. These components, being defined on a spherical shell, can be readily projected onto the orthonormal basis formed by the VSH $\mathbf{Z}_{lm}(\mathbf{k})$, $\mathbf{X}_{lm}(\mathbf{k})$ and $\mathbf{W}_{lm}(\mathbf{k})$ in momentum space, defined in Eq.\ (5) in the Supporting Information. This yields \cite{Fernandez-Corbaton2015}
\begin{equation}
\overset{\circ}{\mathbf{J}}_\omega(\mathbf{k}) = \sum_{lm} a^\omega_{lm} \mathbf{Z}_{lm}(\mathbf{k}) + b^\omega_{lm} \mathbf{X}_{lm}(\mathbf{k}) + c^\omega_{lm} \mathbf{W}_{lm}(\mathbf{k})
\label{Jsp} \:\: ,
\end{equation}
where the $a^\omega_{lm}$, $b^\omega_{lm}$ and $c^\omega_{lm}$ coefficients in this last expression fully describe the radiating source and are therefore called the spherical moments. It is instructive now to study the connection between these multipoles and the Mie coefficients $a^f_{lm}$ and $b^f_{lm}$ in Eq.\ (\ref{Miefield}). We first note that the coefficients $c^\omega_{lm}$, which describe the longitudinal degrees of freedom of $\overset{\circ}{\mathbf{J}}_\omega(\mathbf{k})$, do not radiate outside the source where only transverse fields are present \cite{Fernandez-Corbaton2017}. As such, one can say that $c^f_{lm} = 0$ and, indeed, no $c^f_{lm}$ coefficients are present in the expression of the scattered field in Eq.\ (\ref{Miefield}). As for the other terms, one can use the properties of the Fourier transform to show that $\mathbf{Z}_{lm}(\mathbf{k})$ and $\mathbf{X}_{lm}(\mathbf{k})$ are associated to $\mathbf{N}_{lm}(\mathbf{R})$ and $\mathbf{M}_{lm}(\mathbf{R})$, in such a way that the current is now expressed into the same basis as the scattered field. Indeed, the $a^\omega_{lm}$ $b^\omega_{lm}$ coefficients represent the projection of the current onto the electric and magnetic VSH. As a consequence, one can say that the spherical multipole moments $a^\omega_{lm}$ and $b^\omega_{lm}$ generate the corresponding Mie coefficients $a^f_{lm}$ and $b^f_{lm}$, to which they are indeed related through some simple relations \cite{Alaee2019}. Thanks to this one-to-one correspondence, which we stress holds only in achiral, homogeneous and isotropic systems, $a^\omega_{lm}$ and $b^\omega_{lm}$ represent the \emph{total} current in the source that contributes to the corresponding VSH. As a consequence the field radiated by a spherical multipole, calculated with the help of Eq.\ (\ref{multiE}), will \emph{exactly} match the one projected onto the corresponding VSH calculated with Eq.\ (\ref{Miefield}). For this reason, the spherical multipoles have been sometimes called the "exact" multipoles, but we stress the fact that this nomenclature only stems from their property of exactly matching the predictions of Mie theory, rather than from their ability to exactly describe a scattering system. For instance, in systems lacking spherical symmetry, different multipoles are required to completely characterize the scattering response at a particular order and therefore the irreducible multipoles may show a better convergence since, for example, $a^\omega_{1m}$ does not represent the full electric dipolar response of the scatterer anymore \cite{Evlyukhin2019, Gladyshev2020}. On the other hand, this holds true for spherical particles, where $a^\omega_{1m}$ represents the total electric dipolar response of the scatterer and where there is no need anymore to consider additional toroidal corrections\footnote{Since their radiation properties are similar to those of their parent multipoles, the contributions of the mean square radii of the Cartesian multipoles to the total scattered radiation are also completely accounted for by the spherical multipoles.}. A downside of this is that it is not possible, when employing this formalism, to distinguish different current configurations with similar radiating properties. However, by expanding the spherical Bessel functions $j_l(kr)$ in the definition of the spherical moments (cfr. Eq.\ (9) in the Supporting Information), it is possible to recover the irreducible multipoles and all the relative toroidal corrections \cite{Fernandez-Corbaton2015}. We show this for the case of the spherical electric dipole \cite{Alaee2018}
\begin{equation}
P_\alpha^S = \frac{i}{\omega} \int_V \left\{ J_\alpha j_0(kr) dV + \frac{k^2}{2} \int_V \left[ 3 (\mathbf{r} \cdot \mathbf{J}) r_\alpha - r^2 J_\alpha \right] \frac{j_2(kr)}{(kr)^2} dV \right\} \:\: ,
\end{equation}
of which we write an approximate expression valid for scatterers smaller than the wavelength of the incoming field. This implies $kr << 1$ and, consequently, $j_0(kr) \simeq 1 - (kr)^2 / 6$ and $j_2(kr) \simeq (kr)^2 / 15$, yielding \cite{Alaee2018}
\begin{equation}
P_\alpha^S \simeq \frac{i}{\omega} \int_V J_\alpha dV + \frac{ik}{c} \int_V \frac{1}{10} \left[ (\mathbf{r} \cdot \mathbf{J}) r_\alpha - 2 r^2 J_\alpha \right] dV = P^I_\alpha + \frac{ik}{c} T^P_{1_\alpha} \:\: .
\end{equation}
Clearly, in the dipolar approximation, the spherical electric dipole can be decomposed into the irreducible electric dipole together with its first toroidal correction. Moreover, by considering more terms in the expansion of the spherical Bessel functions, higher order toroidal corrections can also be derived \cite{Fernandez-Corbaton2015}, showing how the spherical multipoles provide a unified description of the effects of both the irreducible and toroidal moments. As a consequence, it is clear now that the toroidal multipoles, rather than being an independent multipole family, are simply a subset of the spherical moments and act as high order corrections to their long-wavelength approximate forms.

As for the Cartesian multipoles, we provide in Table IV of the Supporting Information the definition of the first three orders of spherical multipoles.

\section{Multipolar scattering cross section calculations}\label{S_SCS}

\begin{figure}
\includegraphics[width=0.5\textwidth]{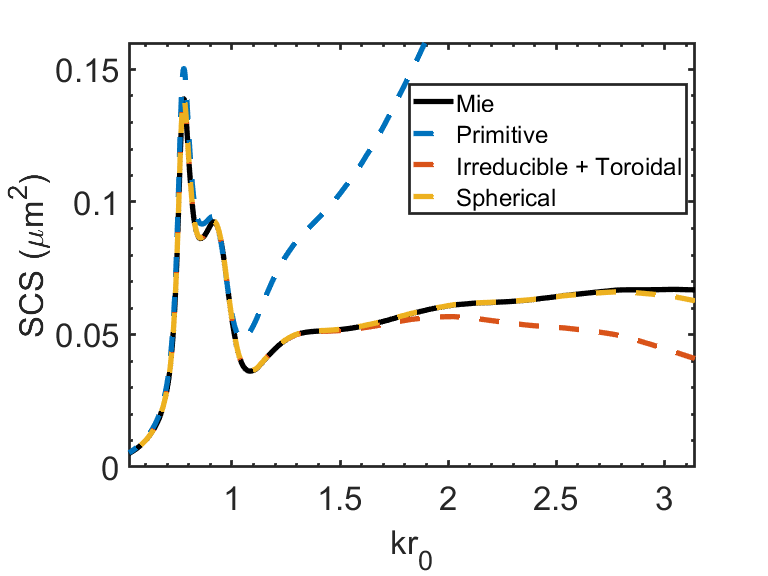}\\
\includegraphics[width=0.5\textwidth]{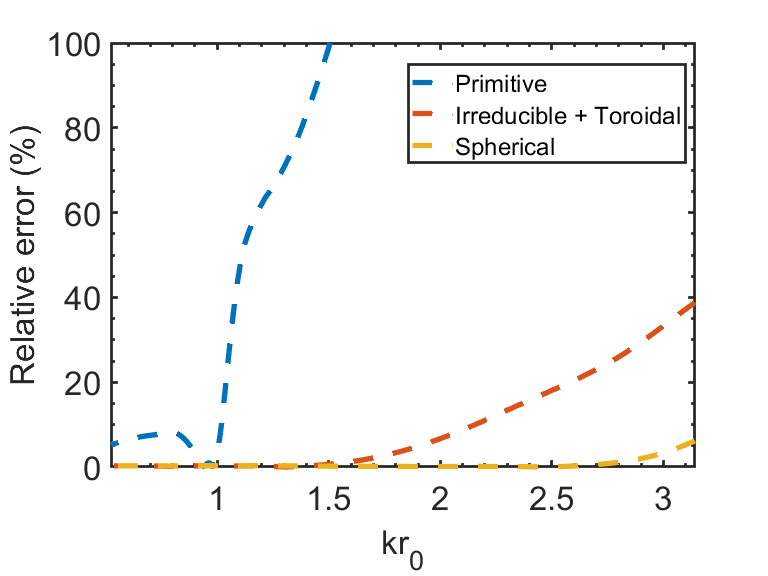}
\caption{Top panel: scattering cross section for an amorphous silicon sphere of radius $r_0 = 100$ nm calculated with the exact Mie theory (full black line) and with the electric and magnetic primitive, irreducible and spherical moments (dashed lines) up to the octupolar order. The irreducible moments are corrected with the first known toroidal multipoles \cite{Gurvitz2019}. Bottom panel: relative error of the multipolar cross sections with respect to the exact Mie solution, calculated as $100 \cdot |SCS_{\text{Mie}} - SCS_{\text{multipoles}}|/SCS_{\text{Mie}}$.}
\label{SCS}
\end{figure}

After reviewing the derivation of these three different multipoles formulations, we can now insert their expressions (see Tables I, II and III in the Supporting Information) into Eq.\ (\ref{multiE}) to compute the different multipolar scattering cross sections and compare them to the exact one given by Mie theory in Eq.\ (\ref{Miefield}). The results are shown in Figure \ref{SCS} for the first three orders of electric and magnetic multipoles as a function of the size parameter $kr_0 = 2\pi r_0 / \lambda$.
\begin{figure*}
\includegraphics{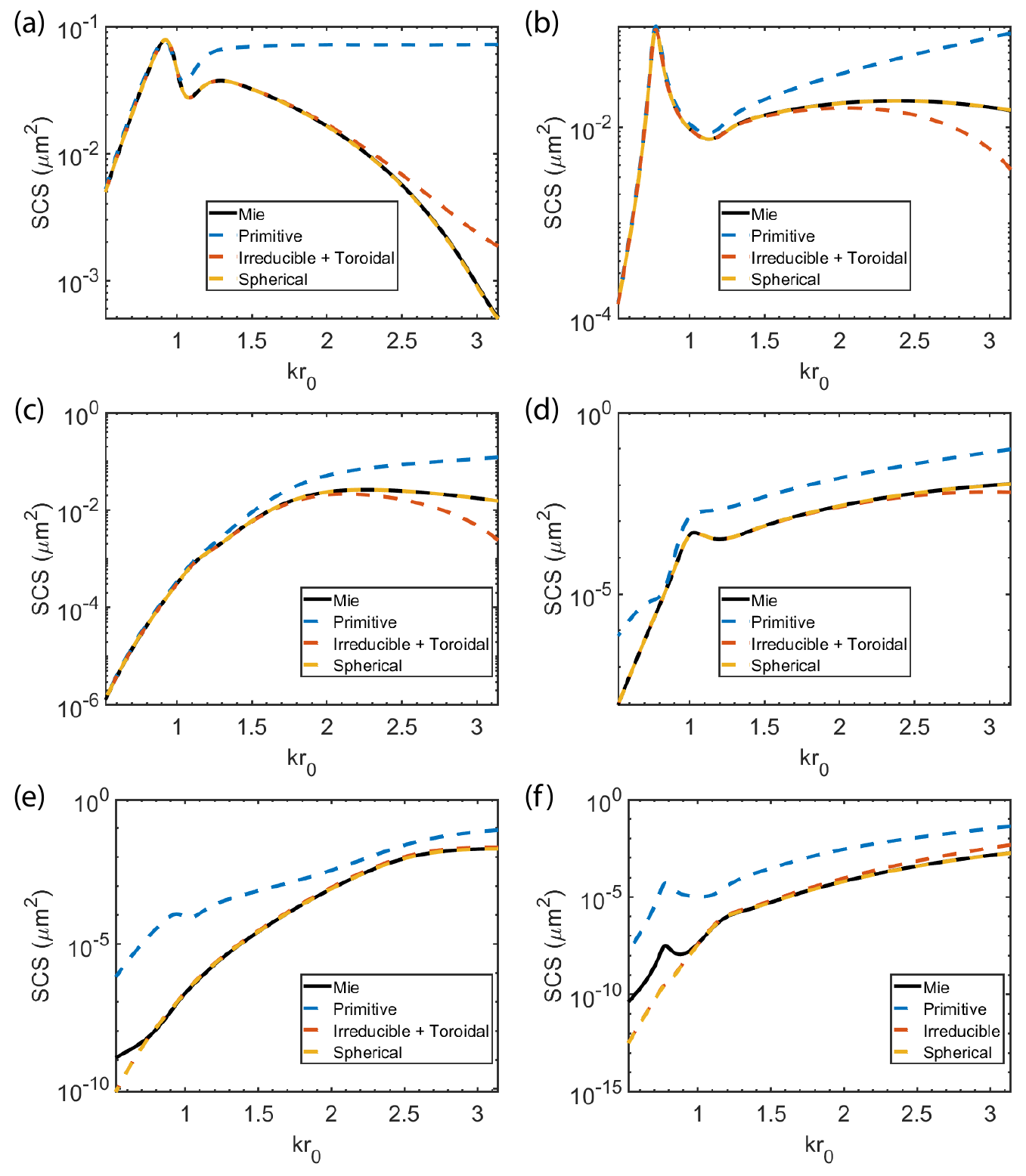}
\caption{First three orders of electric (left) and magnetic (right) cross sections calculated using the vector spherical harmonics decomposition (full black lines) and the three different multipolar formulations (dashed lines) using Eq.\ \ref{multiE}. (a)-(b) Dipolar cross sections. (c)-(d) Quadrupolar cross sections. (e)-(f) Octupolar cross sections.}
\label{fig5}
\end{figure*}
Clearly, the scattering cross section calculated with primitive multipoles cannot approximate the exact one for $kr_0 > 1$, or $\lambda < 2\pi r_0 \simeq 6 r_0$. This confirms the fact that the primitive moments do not provide a good representation of the scattering system. The agreement at $kr_0 < 1$, where the electric and magnetic dipolar responses are dominant (see Figure \ref{VSH} and Figure \ref{fig5}), is simply caused by the fact that the primitive dipole moments are the same as the irreducible ones. On the other hand, the use of irreducible moments, corrected with the first known toroidal multipoles\footnote{The first toroidal moment is known for all multipoles up to the electric octupole and magnetic quadrupole. For the electric dipole, the second toroidal electric moment has also been derived while, to the best of our knowledge, no expressions exist for the toroidal magnetic octupole \cite{Gurvitz2019}. If needed, additional toroidal moments can be derived from the Taylor expansion of the relative spherical multipoles, as already pointed out before.}, is able to improve the agreement at larger size parameters until roughly $kr_0 = 2$, or $\lambda = \pi r_0 \simeq 3 r_0$. In principle, the inclusion of additional toroidal corrections can further improve the results at smaller wavelengths but, as already discussed, the derivation of higher order toroidal multipoles becomes increasingly difficult and is therefore impractical. The best results are clearly obtained when employing the spherical multipoles, which give \emph{exact} results across the entire wavelength range considered. We stress the fact that the small deviation appearing for $kr_0 > 3$ is not due to an imprecise moments definition, but rather to the decision of leaving out the contributions from higher order multipoles, as confirmed by the vector spherical harmonics decomposition of the Mie scattering cross section shown in Figure \ref{VSH}. Note that this deviation reaches a maximum of 6\% even at these high large parameters, while it is virtually zero anywhere else. We further provide in Figure \ref{fig5} the single multipolar cross sections for the different multipole formulations. Again we see that, at least for low order multipoles, primitive Cartesian moments are able to properly describe the system only up to $kr_0 = 1$ where the dipolar responses are dominant, while the corrected irreducible moments can be used up to $kr_0 = 2$. On the other hand, the spherical multipoles provide an exact description of the system as their scattered fields perfectly match the VSH of the corresponding order. The deviation shown at small values of the SCS in Figure \ref{fig5}(e) - (f) can be attributed to numerical errors.

\section{Multipolar optical force calculations}\label{S_force}

\begin{figure}
\includegraphics{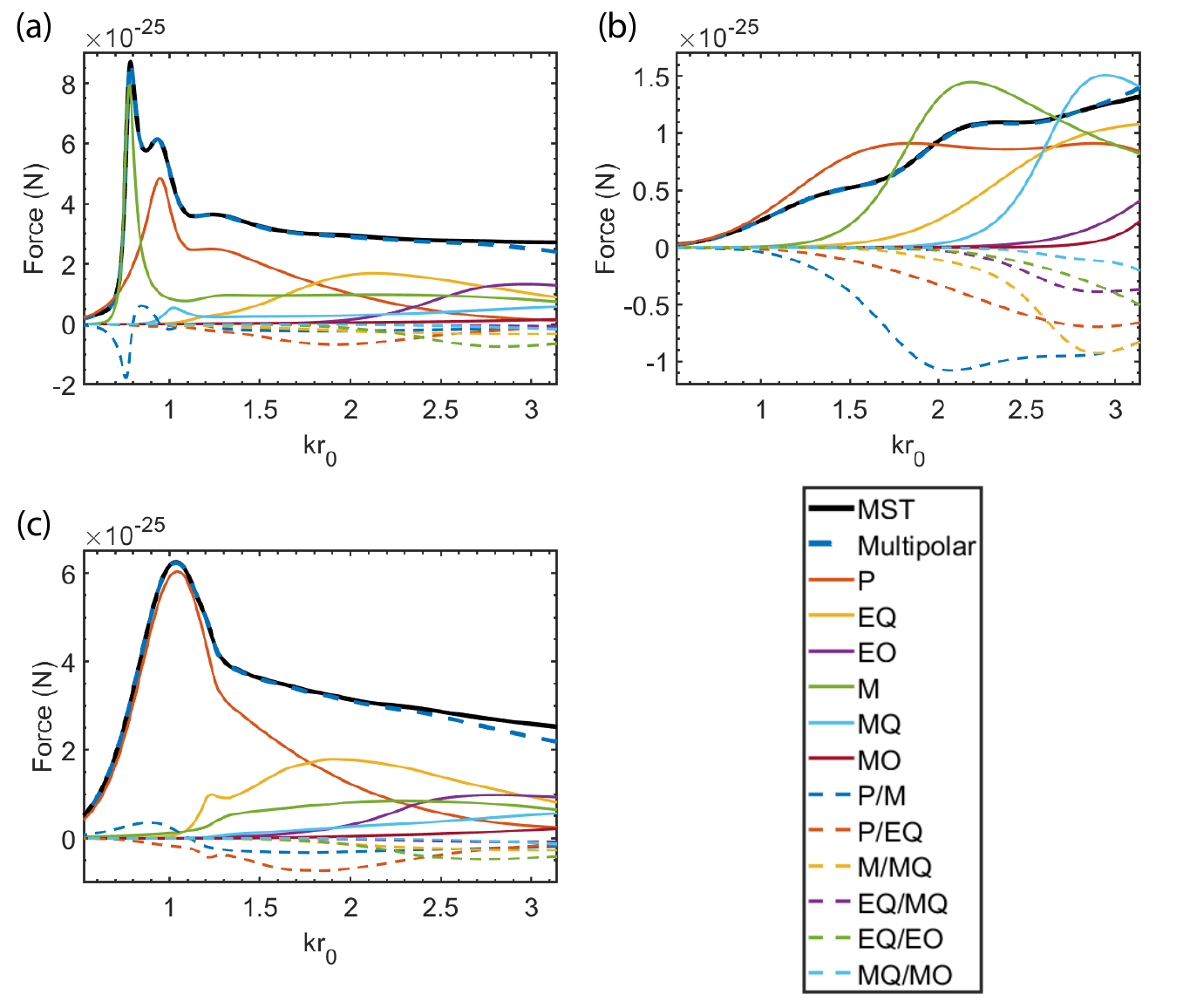}
\caption{Optical force acting on a sphere of radius $r_0 = 100$ nm made of (a) amorphous silicon, (b) glass (n = 1.5) and (c) gold (optical data from \cite{Johnson1972}). The force is calculated both with Maxwell's stress tensor (full black line) and with the spherical moments (thick dashed blue line). Also shown is the decomposition of the multipolar optical force into its components $\mathbf{F}$ (full coloured lines) and $\mathbf{F}_{int}$ (thin dashed lines) using the spherical moments.}
\label{Fmulti}
\end{figure}

After identifying the spherical multipoles as those that better describe the electromagnetic scattering for spherical particles, let us now apply these multipoles to study the optical force exerted on such particles. We compare the exact time-averaged force calculated using Maxwell's stress tensor (MST) \cite{Ji2014} to the multipolar one $\mathbf{F}_{tot} = \mathbf{F} + \mathbf{F}_{int}$ calculated with the first three electric and magnetic spherical moments, which we indicate here with the superscript $S$. This force is the sum of a term $\mathbf{F}$ arising from the interaction between the multipoles and the incoming fields
\begin{eqnarray}
\mathbf{F} = \frac{1}{2} \Re \left( \mathbf{P}^S \cdot \nabla \mathbf{E}^* \right) + \frac{1}{2} \Re \left( \mathbf{M}^S \cdot \nabla \mathbf{B}^* \right) +
\nonumber \\
+ \frac{1}{4} \Re \left( \underline{\underline{\mathbf{EQ}}}^S : \nabla \nabla \mathbf{E}^* \right) + \frac{1}{4} \Re \left( \underline{\underline{\mathbf{MQ}}}^S : \nabla \nabla \mathbf{B}^* \right) +
\label{F} \\
+ \frac{1}{12} \Re \left( \underline{\underline{\underline{\mathbf{EO}}}}^S \dot{:} \nabla \nabla \nabla \mathbf{E}^* \right) + \frac{1}{12} \Re \left( \underline{\underline{\underline{\mathbf{MO}}}}^S \dot{:} \nabla \nabla \nabla \mathbf{B}^* \right)
\nonumber
\end{eqnarray}
and another term $\mathbf{F}_{int}$ representing the interaction between the multipoles themselves
\begin{eqnarray}
\mathbf{F}_{int} = - \frac{k^4}{12 \pi \epsilon_0 c} \Re \left( \mathbf{P}^S \times \mathbf{M}^{S*} \right) - \frac{k^5}{40 \pi \epsilon_0} \Im \left( \mathbf{P}^S \cdot \underline{\underline{\mathbf{EQ}}}^{S*} \right) -
\nonumber \\
- \frac{k^5}{40 \pi \epsilon_0 c^2} \Im \left( \mathbf{M}^S \cdot \underline{\underline{\mathbf{MQ}}}^{S*} \right) - \frac{k^7}{630 \pi \epsilon_0} \Im \left( \underline{\underline{\mathbf{EQ}}}^S : \underline{\underline{\underline{\mathbf{EO}}}}^{S*} \right) -
\nonumber \\
\label{Fint}\\
- \frac{k^7}{630 \pi \epsilon_0 c^2} \Im \left( \underline{\underline{\mathbf{MQ}}}^S : \underline{\underline{\underline{\mathbf{MO}}}}^{S*} \right) - \frac{k^6}{240 \pi \epsilon_0 c} \Re \left( \underline{\mathbf{EQ}}_x^S \times \underline{\mathbf{MQ}}_x^{S*} + \right.
\nonumber \\
\left. + \underline{\mathbf{EQ}}_y^S \times \underline{\mathbf{MQ}}_y^{S*} + \underline{\mathbf{EQ}}_z^S \times \underline{\mathbf{MQ}}_z^{S*} \right) \:\:,
\nonumber
\end{eqnarray}
with $\mathbf{B}$ being the incident magnetic induction field and $^*$ indicating the complex conjugate operation \cite{Chen2011, Jiang2015}. $\mathbf{EQ}_x^S = (EQ_{xx}^S \:\: EQ_{xy}^S \:\: EQ_{xz}^S)$, similarly for $\mathbf{EQ}_y^S$, $\mathbf{EQ}_z^S$ and their magnetic counterparts. We now use these expressions to calculate the force acting on spheres of radius $r_0 = 100$ nm made of different materials under the influence of a field with amplitude of $1 \: V/m$. The results are shown in Figure \ref{Fmulti}, where the $z$-component of the total force, together with its multipolar decomposition, is plotted for a semiconductor, a dielectric or a metallic sphere. This force represents the radiation pressure acting on the particle and is the only force present in the system. We note that, due to the different conventions used to derive Eqs.\ (\ref{F}) - (\ref{Fint}) and the spherical moments, the spherical quadrupoles need to be divided by 3 in order to be used to calculate the multipolar force.
For the force acting on the amorphous silicon sphere, shown in Figure \ref{Fmulti}(a), we see that the magnetic and electric dipolar forces dominate the response of the system roughly until $kr_0 \simeq 0.8$, i.e.\ for $\lambda > 8 r_0$. At shorter wavelengths, until $kr_0 \simeq 2$ or $\lambda \simeq 3r_0$, the magnetic and electric quadrupolar responses need also to be considered to properly characterize the force. For larger size parameters, the electric octupolar force prevails, while the magnetic octupolar response remains negligible. Generally, the smaller the wavelength the more multipoles are excited in the particle, making it harder to single out a dominant contribution from a single multipole. For the dielectric sphere case shown in Figure \ref{Fmulti}(b), the electric rather than the magnetic dipolar force is first generated at small $kr_0$. These dipolar forces dominate the response of the particle until $kr_0 \simeq 1.5$, or for $\lambda > 4 r_0$, when quadrupole moments are then also excited. Moreover, for $kr_0 > 2.5$, i.e. $\lambda < 2.5 r_0$, both the electric and magnetic octupolar forces need to also be included in the calculations. For the metallic particle in Figure \ref{Fmulti}(c) we see that, as expected, the magnetic response is weaker than in the other materials and the electric dipolar force dominates until $kr_0 \simeq 1$, or for $\lambda > 6 r_0$, after which the electric quadrupole and the magnetic dipole come into play. Octupolar forces start to be relevant for $kr_0 > 2$, i.e. $\lambda < 3 r_0$. Similarly to what was reported by Wiscombe for the Mie series \cite{Wiscombe1980}, we provide in Table \ref{table} a guide to judge the correct number of multipoles to consider when performing multipolar optical force calculations. For example, when looking at the amorphous silicon case, one can see that quadrupolar contributions to the optical force emerge at $kr_0 > 0.8$ or $\lambda < 8 r_0$, while octupolar forces start to play a role after $kr_0 > 2$ or $\lambda < 3 r_0$.
\begin{table}
\renewcommand\arraystretch{1.5}
\begin{ruledtabular}
	\begin{tabular}{c|cc|cc}
		 Material & \multicolumn{2}{c}{Quadrupolar} & \multicolumn{2}{c}{Octupolar} \\
			
		\hline
		
		& $kr_0$ & $\lambda$ & $kr_0$ & $\lambda$ \\
		
		\cline{2-5}
			
		Amorphous silicon & 0.8 & $8 r_0$ & 2 & $3 r_0$ \\
			
		Glass & 1.5 & $4 r_0$ & 2.5 & $2.5 r_0$\\
			
		Gold & 1 & $6 r_0$ & 2 & $3 r_0$\\
	\end{tabular}
\end{ruledtabular}
\caption{Approximate threshold values inferred from Figure \ref{Fmulti} for three different materials, for the onset where quadrupolar and octupolar forces need to be included in the multipolar force calculations.}
\label{table}
\end{table}

\section{Conclusions}

We have briefly reviewed the definition and derivation of the primitive and irreducible Cartesian moments and of the spherical multipoles, and used them to perform multipolar scattering cross section and optical force calculations. It is clear from this analysis that, especially when the size of the material system is comparable to the wavelength, the use of Cartesian multipoles should be abandoned in favour of the spherical ones. These clearly outperform the Cartesian ones and provide an \emph{exact} description of the scattering process across the whole wavelength range in achiral, homogeneous and isotropic systems. We also performed a full multipolar decomposition of the optical force up to very large size parameters, providing valuable physical insights into the mechanism of force generation. Finally, we also provided a guide to determine, depending on the working wavelength, how many multipoles orders need to be considered when performing optical force calculations, providing a helpful reference for future works on the topic.

\bibliography{references}

\end{document}


	
\title{Multipole expansions for scattering and optical force calculations beyond the long wavelength approximation - Supporting Information}
	
\author{Marco Riccardi, Andrei Kiselev, Karim Achouri and Olivier J. F. Martin}
\affiliation{Nanophotonics and Metrology Laboratory, Swiss Federal Institute of Technology Lausanne (EPFL), EPFL-STI-NAM, Station 11, CH- 1015 Lausanne, Switzerland}
	
\maketitle

\section{Vector spherical harmonics (VSH)}

\begin{figure*}
\includegraphics[scale=0.8]{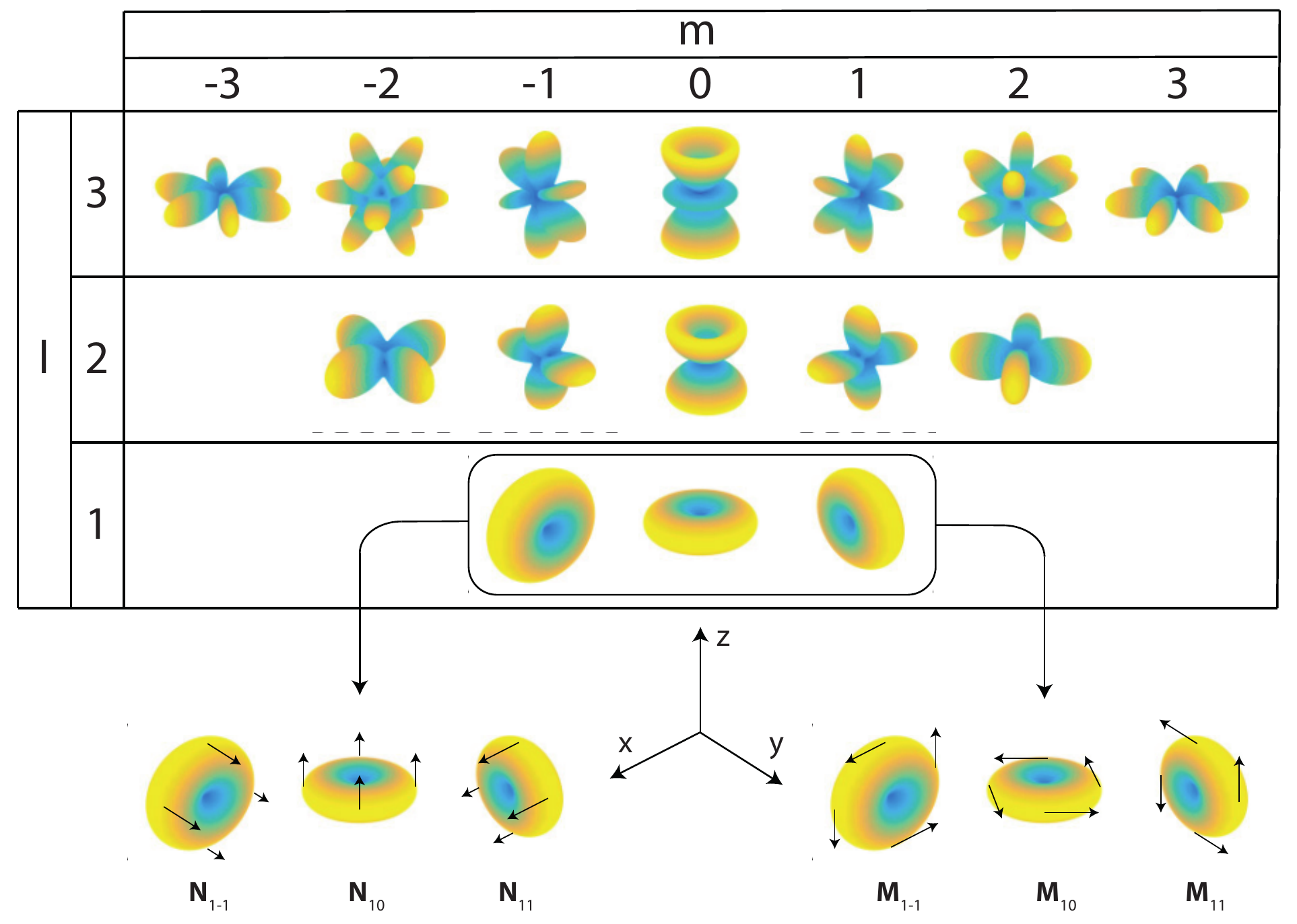}
\caption{Angular part of the first three orders of the vector spherical harmonics. As shown for the $l = 1$ case, the electric and magnetic VSH $\mathbf{N}_{lm}$ and $\mathbf{M}_{lm}$ have the same angular distribution but different field orientations.}
\label{VSH}
\end{figure*}

Given a spherical harmonic in real space
\begin{equation}
Y_{lm}(\theta, \phi) = \sqrt{\frac{2l + 1}{4 \pi} \frac{(l-m)!}{(l + m)!}} P^m_l(cos \theta) e^{i m \phi}
\end{equation}
with $P^m_l(cos \theta)$ being the associated Legendre function, we can define the normalized vector spherical harmonic as \cite{Jackson1998}
\begin{equation}
\mathbf{X}_{lm}(\theta, \phi) = \frac{1}{\sqrt{l (l + 1)}} \mathbf{L} Y_{lm}(\theta, \phi) \:\: .
\end{equation}
The electric and magnetic vector spherical harmonics are then defined, respectively, as \cite{Alaee2019}
\begin{equation}
\mathbf{N}_{lm}(\mathbf{R}) = \frac{1}{k} \nabla \times \mathbf{M}_{lm}(\mathbf{R})
\end{equation}
\begin{equation}
\mathbf{M}_{lm}(\mathbf{R}) = h_l(kR) \mathbf{X}_{lm}(\theta, \phi) \:\: .
\end{equation}
In the above, $\mathbf{L}$ represents the angular momentum operator, $h_l(kR)$ the outgoing spherical Hankel functions (Hankel functions of the first kind) and the other symbols have got the same meaning as in the main text. Figure \ref{VSH} shows the angular part of the first few VSH.

By exploiting the properties of the Fourier transform \cite{Alaee2019}, it is possible to derive the VSH in momentum space:
\begin{eqnarray}
	\mathbf{Z}_{lm}(\mathbf{k}) = i \mathbf{k} \times \mathbf{X}_{lm}(\mathbf{k})
	\nonumber \\
	\mathbf{X}_{lm}(\mathbf{k}) = \frac{1}{\sqrt{l(l+1)}} \mathbf{L} Y_{lm}(\mathbf{k})
	\label{basisF}\\
	\mathbf{W}_{lm}(\mathbf{k}) = \mathbf{k} Y_{lm}(\mathbf{k})
	\nonumber \:\: ,
\end{eqnarray}
where $Y_{lm}(\mathbf{k})$ is the Fourier transform of $Y_{lm}(\theta, \phi)$.
	
\section{Cartesian multipoles}

The general expressions of the primitive Cartesian electric and magnetic multipoles are \cite{1386}
\begin{equation}
P^{P(l)}_{i_1 \dots i_l}(\tau)= \int \rho(\mathbf{r}, \tau) \prod_{k = 1}^{l} r_{i_k} d\mathbf{r}
\end{equation}
\begin{equation}
M^{P(l)}_{i_1 \dots i_l}(\tau)= \frac{l}{l+1} \int \left[ \mathbf{r} \times \mathbf{J}(\mathbf{r}, \tau) \right] \prod_{k = 1}^{l-1} r_{i_k} d\mathbf{r} \:\: ,
\end{equation}
where $\rho(\mathbf{r}, \tau)$ represents the charge density in the scatterer, and the other variables have the same meaning as in the main text. The expressions of first three orders of primitive multipoles are explicitly given in Table \ref{tab1}.
\begin{table*}
\caption{First three orders of primitive multipoles \cite{Gurvitz2019}.}
\renewcommand\arraystretch{1.5}
\begin{ruledtabular}
	\begin{tabular}{ccc}
		l & Primitive electric multipoles & Primitive magnetic multipoles \\
			
		\hline
			
		1 & $\left( \mathbf{P}^P \right)_{\alpha} = P_\alpha^P = \frac{i}{\omega} \int_V J_\alpha dV$ & $\left( \mathbf{M}^P \right)_{\alpha} = M_\alpha^P = \frac{1}{2} \int_V \left( \mathbf{r} \times \mathbf{J} \right)_\alpha dV$ \\
		
		2 & $\left( \underline{\underline{\mathbf{P}}}^P \right)_{\alpha \beta} = EQ_{\alpha \beta}^P = \frac{i}{\omega} \int_V \left( r_\alpha J_\beta + r_\beta J_\alpha \right) dV$ & $\left( \underline{\underline{\mathbf{M}}}^P \right)_{\alpha \beta} = MQ_{\alpha \beta}^P = \frac{2}{3} \int_V r_\alpha \left( \mathbf{r} \times \mathbf{J} \right)_\beta dV$ \\
			
		3 & $\left( \underline{\underline{\underline{\mathbf{P}}}}^P \right)_{\alpha \beta \gamma} = EO_{\alpha \beta \gamma}^P = \frac{i}{\omega} \int_V \left( r_\alpha r_\beta J_\gamma + r_\gamma r_\alpha J_\beta + r_\beta r_\gamma J_\alpha \right) dV$ & $\left( \underline{\underline{\underline{\mathbf{M}}}}^P \right)_{\alpha \beta \gamma} = MO_{\alpha \beta \gamma}^P = \frac{3}{4} \int_V r_\gamma r_\beta \left( \mathbf{r} \times \mathbf{J} \right)_\alpha dV$ \\
	\end{tabular}
\end{ruledtabular}
\label{tab1}
\end{table*}

Table \ref{tab2} displays the first three orders of irreducible Cartesian moments. In the definition of the irreducible magnetic octupole, the symmetric part of the primitive magnetic octupole $\underline{\underline{\underline{\mathbf{O}}}}^{sym}$ appears, which is defined as
\begin{equation}
O_{\alpha \beta \gamma}^{sym} = \frac{1}{3} (MO_{\alpha \beta \gamma}^P + MO_{\beta \alpha \gamma}^P + MO_{\gamma \alpha \beta}^P) \:\: .
\end{equation}
We point out that the expression of the irreducible magnetic octupole provided by Gurvitz \emph{et al.} in the main text is not correct \cite{Gurvitz2019}. However, correct expressions can be found in the Supporting Information of their work.
\begin{table*}	
	\caption{First three orders of irreducible multipoles, where the Einstein's summation convention is implied \cite{Gurvitz2019}.}
	\renewcommand\arraystretch{1.5}
	\begin{ruledtabular}
		\begin{tabular}{ccc}
			l & Irreducible electric multipoles & Irreducible magnetic multipoles \\
			
			\hline
			
			1 & $P_\alpha^I = P_\alpha^P = \frac{i}{\omega} \int_V J_\alpha dV$ & $M_\alpha^I = M_\alpha^P = \frac{1}{2} \int_V \left( \mathbf{r} \times \mathbf{J} \right)_\alpha dV$ \\
			
			2 & $EQ_{\alpha \beta}^I = \frac{i}{\omega} \int_V \left[ r_\alpha J_\beta + r_\beta J_\alpha - \frac{2}{3} \delta_{\alpha \beta} \left( \mathbf{r} \cdot \mathbf{J} \right) \right] dV$ & $MQ_{\alpha \beta}^I = \frac{1}{3} \int_V \left[ r_\alpha \left( \mathbf{r} \times \mathbf{J} \right)_\beta + r_\beta \left( \mathbf{r} \times \mathbf{J} \right)_\alpha \right] dV$ \\
			
			3 & $EO_{\alpha \beta \gamma}^I = EO^P_{\alpha \beta \gamma} - \frac{1}{5} \left( \delta_{\alpha \beta} EO^P_{\delta \delta \gamma} + \delta_{\alpha \gamma} EO^P_{\delta \delta \beta} + \delta_{\beta \gamma} EO^P_{\delta \delta \alpha} \right)$ & $MO_{\alpha \beta \gamma}^I = O^{sym}_{\alpha \beta \gamma} - \frac{1}{5} \left( \delta_{\alpha \beta} O^{sym}_{\delta \delta \gamma} + \delta_{\alpha \gamma} O^{sym}_{\delta \delta \beta} + \delta_{\beta \gamma} O^{sym}_{\delta \delta \alpha} \right)$ \\
			
		\end{tabular}
	\end{ruledtabular}
	\label{tab2}
\end{table*}

Table \ref{tab3} provides the first known three orders of toroidal moments \cite{Gurvitz2019}.
\begin{table*}	
	\caption{First known three orders of toroidal multipoles, where the Einstein's summation convention is implied \cite{Gurvitz2019}.}
	\renewcommand\arraystretch{1.5}
	\begin{ruledtabular}
		\begin{tabular}{ccc}
			l & Toroidal electric multipoles & Toroidal magnetic multipoles \\
			
			\hline
			
			\multirow{2}*{1} & $T^P_{1_\alpha} = \frac{1}{10} \int_V \left[ (\mathbf{r} \cdot \mathbf{J}) r_\alpha - 2 r^2 J_\alpha \right] dV$ & $T^M_{1_\alpha} = \frac{i \omega}{20} \int_V \left[ r^2 (\mathbf{r} \times \mathbf{J})_\alpha \right] dV$ \\
			
			& $T^P_{2_\alpha} = \frac{1}{280} \int_V \left[ 3 r^4 J_\alpha - 2 r^2 (\mathbf{r} \cdot \mathbf{J}) r_\alpha \right] dV$ & \\
			
			\hline
			
			2 & $T^{EQ}_{1_{\alpha \beta}} = \frac{1}{42} \int_V \left[ 4 (\mathbf{r} \cdot \mathbf{J}) r_\alpha r_\beta + 2 r^2 (\mathbf{r} \cdot \mathbf{J}) \delta_{\alpha \beta} - 5 r^2 (r_\alpha J_\beta + r_\beta J_\alpha) \right] dV$ & $T^{MQ}_{1_{\alpha \beta}} = \frac{i \omega}{42} \int_V r^2 \left[ r_\alpha (\mathbf{r} \times \mathbf{J})_\beta + r_\beta (\mathbf{r} \times \mathbf{J})_\alpha \right] dV$ \\
			
			\hline
			
			\multirow{2}*{3} & $T^{EO}_{1_{\alpha \beta \gamma}} = \frac{1}{300} \int_V \left\{ 35 (\mathbf{r} \cdot \mathbf{J}) r_\alpha r_\beta r_\gamma - 20 r^2 (r_\alpha r_\beta J_\gamma + r_\gamma r_\alpha J_\beta + r_\beta r_\gamma J_\alpha) + \right. $ &  \\
			
			& $ \left. + (\delta_{\alpha \beta} \delta_{\gamma \delta} + \delta_{\gamma \alpha} \delta_{\beta \delta} + \delta_{\beta \gamma} \delta_{\alpha \delta}) \left[ r^2 (\mathbf{r} \cdot \mathbf{J}) r_\delta + 4 r^4 J_\delta \right] \right\} dV $ & \\
			
		\end{tabular}
	\end{ruledtabular}
	\label{tab3}
\end{table*}

\section{Spherical multipoles}

The general expression of a spherical multipole $q^\omega_{lm} = \{a^\omega_{lm}, b^\omega_{lm}\}$ in spherical coordinates is \cite{Alaee2019}
\begin{equation}
q^\omega_{lm} = \frac{4\pi}{\sqrt{(2\pi)^3}} \sum_s \sum_{u = -s}^{u = s} (-i)^s \left[ \int d\mathbf{k} \mathbf{Q}^+_{lm} Y_{su}(\mathbf{k}) \int d\mathbf{r} \mathbf{J}_\omega(\mathbf{r}) Y^+_{su}(\mathbf{r}) j_s(kr) \right] \:\: ,
\end{equation}
where $\mathbf{Q}_{lm} = \{\mathbf{Z}_{lm}, \mathbf{X}_{lm}\}$, $j_s(kr)$ is the spherical Bessel function of order $s$ and the other symbols have the same meaning as in the main text. The expressions of the first three orders of spherical multipoles, expressed in Cartesian coordinates, are provided in Table \ref{tab4}. Formulas to convert a spherical multipole $q^\omega_{lm}$ in Cartesian coordinates can be found in \cite{Alaee2018, Mun2020}.
\begin{table*}	
\caption{First three orders of the spherical multipoles in Cartesian coordinates \cite{Alaee2018, Evlyukhin2019, Mun2020}.}	\renewcommand\arraystretch{1.5}
\begin{ruledtabular}
	\begin{tabular}{ccc}
		l & Spherical electric multipoles & Spherical magnetic multipoles \\
			
		\hline
		
		1 & $P_\alpha^S = \frac{i}{\omega} \int_V \left\{ J_\alpha j_0(kr) dV + \frac{k^2}{2} \int_V \left[ 3 (\mathbf{r} \cdot \mathbf{J}) r_\alpha - r^2 J_\alpha \right] \frac{j_2(kr)}{(kr)^2} dV \right\}$ & $M_\alpha^S = \frac{3}{2} \int_V \left( \mathbf{r} \times \mathbf{J} \right)_\alpha \frac{j_1(kr)}{kr} dV$ \\
		
		\hline
			
		\multirow{2}*{2} & $EQ_{\alpha \beta}^S = \frac{3i}{\omega} \int_V \left\{ \left[ 3 (r_\alpha J_\beta + r_\beta J_\alpha) - 2 \delta_{\alpha \beta} (\mathbf{r} \cdot \mathbf{J}) \right] \frac{j_1(kr)}{kr} dV + \right.$ & \multirow{2}*{$MQ_{\alpha \beta}^S = 5 \int_V \left[ r_\alpha \left( \mathbf{r} \times \mathbf{J} \right)_\beta + r_\beta \left( \mathbf{r} \times \mathbf{J} \right)_\alpha \right] \frac{j_2(kr)}{(kr)^2} dV$} \\
		
		& $\left. + 2k^2 \int_V \left[ 5 r_\alpha r_\beta (\mathbf{r} \cdot \mathbf{J}) - (r_\alpha J_\beta + r_\beta J_\alpha) r^2 - r^2 \delta_{\alpha \beta} (\mathbf{r} \cdot \mathbf{J}) \right] \frac{j_3(kr)}{(kr)^3} dV \right\}$ & \\
		
		\hline
			
		\multirow{5}*{3} & $EO_{\alpha \beta \gamma}^S = \frac{15i}{\omega} \int_V \left\{ r_\alpha r_\beta J_\gamma + r_\alpha r_\gamma J_\beta + r_\beta r_\gamma J_\alpha  - \right.$ &  \\
		
		&$-\frac{1}{5} \left[ \delta_{\beta \gamma} \left( 2 r_\alpha (\mathbf{r} \cdot \mathbf{J}) + r^2 J_\alpha \right) + \delta_{\alpha \gamma} \left( 2 r_\beta (\mathbf{r} \cdot \mathbf{J}) + r^2 J_\beta \right) + \right.$& $MO_{\alpha \beta \gamma}^S = \frac{105}{4} \int_V \left\{ r_\alpha r_\beta \left( \mathbf{r} \times \mathbf{J} \right)_\gamma + r_\alpha r_\gamma \left( \mathbf{r} \times \mathbf{J} \right)_\beta + \right.$ \\

		& $\left. \left. + \delta_{\alpha \beta} \left( 2 r_\gamma (\mathbf{r} \cdot \mathbf{J}) + r^2 J_\gamma \right) \right] \frac{j_2(kr)}{(kr)^2} dV \right\} + \frac{45k^2i}{4 \omega} \int_V \left\{ 7 r_\alpha r_\beta r_\gamma (\mathbf{r} \cdot \mathbf{J}) - \right.$ & $+ r_\beta r_\gamma \left( \mathbf{r} \times \mathbf{J} \right)_\alpha - \frac{r^2}{5} \left[ \delta_{\beta \gamma} (\mathbf{r} \times \mathbf{J})_\alpha + \right.$ \\
		
		& $- r^2 (r_\alpha r_\beta J_\gamma + r_\alpha r_\gamma J_\beta + r_\beta r_\gamma J_\alpha) + \frac{1}{5} r^2 \delta_{\beta \gamma} \left( r^2  J_\alpha - 5 r_\alpha (\mathbf{r} \cdot \mathbf{J}) \right) +$ & $\left. \left.+ \delta_{\alpha \gamma} (\mathbf{r} \times \mathbf{J})_\beta + \delta_{\alpha \beta} (\mathbf{r} \times \mathbf{J})_\gamma \right] \frac{j_3(kr)}{(kr)^3} dV \right\}$ \\
		
		& $ \left. + \frac{1}{5} r^2 \delta_{\alpha \gamma} \left( r^2  J_\beta - 5 r_\beta (\mathbf{r} \cdot \mathbf{J}) \right)  + \frac{1}{5} r^2 \delta_{\alpha \beta} \left( r^2  J_\gamma - 5 r_\gamma (\mathbf{r} \cdot \mathbf{J}) \right) \frac{j_4(kr)}{(kr)^4} dV \right\}$ & \\
			
	\end{tabular}
\end{ruledtabular}
\label{tab4}
\end{table*}

\bibliography{references}